\documentclass[3p,times,twocolumn]{elsarticle}
 \biboptions{comma,sort&compress}
 
\usepackage{graphicx}
\usepackage{ecrc}

 \usepackage{hyperref}
  \usepackage{color}

\volume{00}

\firstpage{1}

\journalname{Nuclear and Particle Physics Proceedings}

\runauth{}

\jid{nppp}

\jnltitlelogo{Nuclear and Particle Physics Proceedings}

\usepackage{amssymb}

\usepackage[figuresright]{rotating}

\begin{document}

\begin{frontmatter}

\title{The Relation between the Perturbative QCD Scale $\Lambda_s$ and Hadronic Masses from Light-Front Holography}
 \cortext[cor0]{Talk given at 18th International Conference on Quantum Chromodynamics (QCD 15,  30th anniversary),  29 June - 3 July 2015,  Montpellier, France}
 \cortext[cor1]{Preprint: JLAB-PHY-15-2135; SLAC-PUB-16385}
 \author[label1]{A. Deur\fnref{fn1}}
\ead{deurpam@jlab.org}
\fntext[fn1]{Speaker, Corresponding author.}
\address[label1]{Thomas Jefferson National Accelerator Facility, Newport
News, VA 23606, USA}
 \author[label2]{S. J. Brodsky}
 \address[label2]{SLAC National Accelerator Laboratory, Stanford University,
Stanford, California 94309, USA}
 \author[label3]{G. F. de Teramond}
\address[label3]{Universidad de Costa Rica, San Jos\'e, Costa Rica}

\pagestyle{myheadings}
\markright{ }
\begin{abstract}
QCD is well understood at short distances where perturbative calculations are feasible. 
Establishing an explicit analytic connection between the short-distance regime and the large-distance physics of 
quark confinement has been a long-sought goal. A major challenge is to relate the scale $\Lambda_{s}$ underlying the evolution of the 
QCD coupling in the perturbative regime to the masses of hadrons. We show here how new theoretical 
insights into the behavior of QCD at large distances leads to such a relation. The resulting 
prediction for $\Lambda_{s}$ in the $\overline{MS}$ scheme agrees well with experimental measurements.  Conversely,  the relation
can be used to predict the masses of hadrons composed of light quarks with the measured
value of $\Lambda_{s}$ as the sole parameter.  We  also use  ``light-front holography"  to determine the analytic  form of $\alpha_s(Q^2)$ at small $Q^2.$  

\end{abstract}

\end{frontmatter}

\section{Introduction}

The masses of hadrons composed of light quarks such as the  proton and $\rho$ meson are understood to originate from the energy of the confining interactions of QCD; 
however, it is unclear why the typical hadron mass scale is of order $1$ GeV.   One would expect this mass scale to 
be explicitly present in the QCD Lagrangian.  However, the only scale in ${\cal L}_{QCD}$ are the quark masses, which 
for the up and down quarks, are evidently too small to be relevant: $m_q \sim 10^{-3}$ GeV.  A relevant mass scale, 
$\Lambda_s$, however, does exist. It controls the strength of the coupling of quarks when they interact at short 
distances. Its precise definition emerges  when one renormalizes the QCD coupling $\alpha_s(Q^2)$.  The results presented in this talk can be discussed in any choice of renormalization 
scheme, but we will use here the  value of $\Lambda_s$ defined in the  ${\overline{MS}}$ 
(modified minimal subtraction) renormalization scheme. The value of the parameter 
$\Lambda_s =\Lambda_{\overline{MS}}$  can be determined to high precision from experimental 
measurements of  high-energy, short-distance  processes where the strength of QCD  is small because 
of asymptotic freedom~\cite{Gross:1973id, Politzer:1973fx} and perturbative QCD (pQCD) is thus 
applicable. One long-sought goal in 
QCD is to find an explicit relation between the 
hadron masses and  $\Lambda_s$. 

In this talk we present such relation~\cite{Deur:2014qfa},
which leads to the prediction of the value of  
$\Lambda_s$ from a hadronic mass.  Conversely, one can obtain the hadronic spectrum using 
$\Lambda_s$. To establish 
this relation, we use the QCD effective coupling $\alpha_s$ computed at 
small-distance using pQCD and at long distance using the formalism of 
QCD on the Light Front which allows, under reasonable approximations, non-perturbative
calculations.  We will also use  ``light-front holography"  to determine   the precise form of $\alpha_s(Q^2)$ at small $Q^2.  $
The  small and large distance regimes of QCD overlap, a phenomenon related to 
``quark-hadron duality"~\cite{Bloom:1970xb}. This allows us to match the two descriptions and 
obtain the behavior of $\alpha_s(Q^2)$ at any scale. This in turn leads to an analytical relation between 
$\Lambda_s$  and hadron masses.

\section{Light-Front  QCD}

The  light-front (LF)
 quantization procedure is based on the ``Front Form"  invented by Dirac~\cite{Dirac}, where
the time evolution variable is $\tau = t+z/c$;  {\it i.e,} time along the light-front.
The resulting  LF Hamiltonian and its  eigensolutions are Lorentz frame-independent~\cite{LF-QCD}.

One can derive a one-dimensional  ``light-front Schr\"odinger equation"  (LFSE)  in QCD describing the valence Fock state of  color-singlet $q \bar q$ mesons for light quarks, analogous to the
Schr\"odinger equation describing hydrogenic atoms in QED~\cite{deTeramond:2008ht}.   Unlike the QED form, the LFSE is relativistic and frame-independent. 
The radial variable for the LFSE, the invariant separation between the $q$ and $\bar q$  is $\zeta = b_\perp \sqrt{x(1-x)}$, where $\zeta^2 $ is conjugate to the LF kinetic energy $k^2_\perp / (x(1-x))$, the invariant mass squared of the $q \bar  q$.   Here $b_\perp$ is the transverse impact parameter and $x$ is the LF momentum fraction $x= k^+ / P^+ = (k^0 +k^z)/ (P^0 + P^z)$. The  LFSE  incorporates color confinement and other essential spectroscopic and dynamical features of hadron physics, including a massless pion for zero quark mass and linear Regge trajectories with the same slope  in the radial quantum number $n$ and internal  orbital angular momentum $L$.

The form of the LF potential $V(\zeta^2)$ entering the LFSE  -its sole unspecified component- 
becomes uniquely  determined  as a harmonic oscillator $V(\zeta^2) = \kappa^4 \zeta^2$
 when one extends the formalism  of de Alfaro, Fubini and Furlan (dAFF)~\cite{deAlfaro:1976je} to light-front Hamiltonian theory~\cite{Brodsky:2013ar}. This  discovery by dAFF,  in the  context of $1+1$  quantum mechanics, allows for the emergence  in the  theory
of a mass scale  $\kappa$ without it appearing explicitly in the Lagrangian. That is, enforcing the conformal symmetry of QCD fully determines the confinement potential $\kappa^4 \zeta^2$  in the LFSE underlying the hadron spectrum.


The harmonic oscillator  form of the LF potential corresponds
to a linear potential for bound states of heavy quarks in
the usually employed instant-form~\cite{the:Trawinski 2014}.  This  links  
 a semi-classical approximation to light-front QCD, based on the underlying conformality of QCD in the 
 limit of zero quark masses, to   lattice gauge theory and
other approaches to heavy quark effective theory. 
The parameter $\kappa$ is obtained from a hadron mass, e.g. 
$\kappa= M_\rho/ \sqrt{2}$~\cite{Brodsky:2014yha}. 
This provides a  rather model-independent tractable formalism for addressing the non-perturbative QCD bound-state problem  at leading order.

\section{ LF holography}

We have stressed the importance of the conformal symmetry for QCD. The conformal group  in four dimensions is geometrically represented by the five-dimensional AdS$_5$ space.
It is holographically dual to 3+1  spacetime using light-front time $\tau$. 
In this correspondence   the LF variable
$\zeta$ can be identified with the fifth AdS dimension. For hadrons probed  at short distances 
$\zeta  \sim 1/Q^2$,
  with $Q^2$ the 4-momentum squared exchanged between the hadron and a beam particle.

Remarkably, the same confining LF potential $V(\zeta^2) = \kappa^4 \zeta^2$  and the same LFSE for mesons of arbitrary spin $J$ can be derived~\cite{deTeramond:2013it}
from the ``soft-wall model"~\cite{Karch:2006pv} modification of AdS$_5$ space assuming the specific ``dilaton profile" $e^{+\kappa^2 z^2}$.  Using LF Holography,  one  can identifies the fifth dimension coordinate $z$ of AdS$_5$ space with the light-front coordinate $\zeta$.  
This correspondence, often  called AdS/QCD,  is  well established: there exists a one-to-one mapping between 
LF and AdS wavefunctions. Furthermore, the expressions for  the electromagnetic and gravitational form factors of hadrons in AdS$_5$ are the same as the Drell-Yan West formula 
in 3+1 space using LF time~\cite{Brodsky:2006uqa}. All in all, the AdS/QCD correspondence provides an 
excellent description of hadrons of arbitrary spin, incorporating many of
 observed spectroscopic and dynamical features~\cite{Brodsky:2014yha, deTeramond:2013it, Brodsky:2006uqa, deTeramond:2008ht, Brodsky:2013ar}.

\section{ Determining $\alpha_s$ at all scales}

\subsection{ $\alpha_s$ at small $Q^2$ from LF Holography}
One can derive the explicit form for  $\alpha_s(Q^2)$ from LF-QCD using AdS/QCD. 
As noted above, the forces that bind quarks are related  in AdS/QCD to the modification of the AdS space 
curvature,  the dilaton profile  $e^{+\kappa^2 z^2}$ encoding confinement  dynamics~\cite{Brodsky:2013ar}. This modification of 
the AdS geometry is constrained by the form of the potential dictated by the dAFF mechanism.
The same constraint also prescribes the form of $\alpha_s$ at small $Q^2$.  

In pQCD, the effective coupling $\alpha_s(Q^2)$ 
is defined at high  $Q^2$ by folding short-distance quantum effects into its evolution.  
Analogously,  as we will show, the $Q^2$-dependence of the AdS/QCD effective coupling stems from the effects of the 
large-distance forces folded into the coupling constant~\cite{Brodsky:2010ur}.

To determine $\alpha_{g_1}^{AdS}$, consider first the AdS action. It has the same form as
General Relativity's action:
\begin{equation}
S \propto \int d^{4}x \sqrt{det(g_{\mu\nu})} {R \over G_{N}},
\end{equation}
but with $R$, the Ricci scalar, and $G_{N}$, Newton's constant, replaced by their QCD-analogs.
Thus $\sqrt{R}$ is replaced by the gluon field $F$, $\sqrt{G_{N}}$ corresponds to the gauge coupling $g_{AdS}$, 
and the metric determinant is  $\sqrt{det(g^{AdS}_{\mu\nu})}~e^{\kappa^2 z^2}$,
which includes the $e^{\kappa^{2}z^{2}}$ dilaton profile.  
The 5-dimensional AdS action is thus:
\begin{equation}
S=\frac{1}{4}\intop d^5 x  \sqrt{det(g^{AdS}_{\mu \nu})}~ e^{\kappa^{2}z^{2}} \frac{1}{g_{AdS}^{2}} F^{2}.
\end{equation}
In pQCD, $\alpha_{s}\equiv {g_s^{2}/ 4\pi}$ acquires its $Q^2$-dependence from short-distance
quantum effects.   Similarly,  the initially constant AdS coupling $\alpha_{AdS}\equiv g_{AdS}^{2}/4\pi$ is redefined to
absorb the effects of the AdS deformation:
$g_{AdS}^{2}\rightarrow g_{AdS}^{2}~e^{\kappa^{2}z^{2}}$. 
Transforming to momentum space yields~\cite{Brodsky:2010ur}
\begin{equation} \label{alphaAdS}
\alpha^{AdS}_{g_1}(Q^2) = \pi  \exp{\left(-Q^2\over 4 \kappa^2\right)}.
\end{equation}
Here, $\alpha_{g_1}^{AdS}(Q^2)$,   the effective charge in the $g_1$-scheme 
defined from the Bjorken sum rule~\cite{Bjorken:1966jh, Bjorken:1969mm}, is 
normalized to $\pi$ at $Q^2=0$ due to fulfill straightforward kinematical 
constraints~\cite{Deur:2005cf, Deur:2008rf}. This coupling can serve as the 
QCD-analog of the Gell-Mann-Low coupling $\alpha(Q^2)$ of QED~\cite{Brodsky:2010ur}. 
Since LF-holography neglects quantum effects, the short-distance phenomena  which lead to the pQCD evolution  of the running 
of pQCD coupling  are not  incorporated in $\alpha_{AdS}$.  Indeed,  the Gaussian form of Eq. (\ref{alphaAdS}) falls much faster than the pQCD prediction at large $Q^2$.
 
\subsection{Behavior of $\alpha_s$ at large $Q^2$}
The large $Q^2$-dependence of $\alpha_{s}(Q^2)$ is well known~\cite{PDG:2014}. Its evolution is given by 
the QCD renormalization group equation where the logarithmic derivative of the coupling defines the $\beta$ function.   
If $\alpha_s$ is small, one can use the perturbative expansion:
\begin{eqnarray} \label{beta}
Q^2{d \alpha_{s}}/{dQ^2} =
-(\beta_0 \alpha_s^2 + \beta_1 \alpha_s^3+ \beta_2 \alpha_s^4 +  \cdots).
\end{eqnarray}
The $\beta_i$ for $i  \ge 2$ are scheme-dependent  and are known up to order $\beta_3$  in the $\overline{MS}$ renormalization scheme~\cite{PDG:2014}.
Eq. (\ref{beta}) thus yields $\alpha_{\overline{MS}}(Q^2)$ at high $Q^2$.
In addition, 
$\alpha_{g_1}^{pQCD}(Q^2)$ can be expressed as a perturbative expansion in 
$\alpha_{\overline{MS}}(Q^2)$~\cite{Bjorken:1966jh, Bjorken:1969mm}.
Thus, pQCD predicts the form of  $\alpha_{g_1}(Q^2)$ at large $Q^2$:
\begin{eqnarray}  
\label{eq: alpha_g1 from Bj SR}
\alpha_{g_{1}}^{pQCD}(Q^2)  = \pi \Big[{\alpha_{\overline{MS}}}/{\pi} +  
a_1\left({\alpha_{\overline{MS}}}/{\pi}\right)^{2} +   \nonumber
\\
a_2\left({\alpha_{\overline{MS}}}/{\pi}\right)^{3} 
+ a_3\left({\alpha_{\overline{MS}}}/{\pi}\right)^{4} + \cdots \Big].
\end{eqnarray} 
The coefficients $a_i$ are known up to order $a_3$~\cite{Baikov:2010je}.

\begin{figure}[h]
\centering
\includegraphics[width=.42\textwidth]{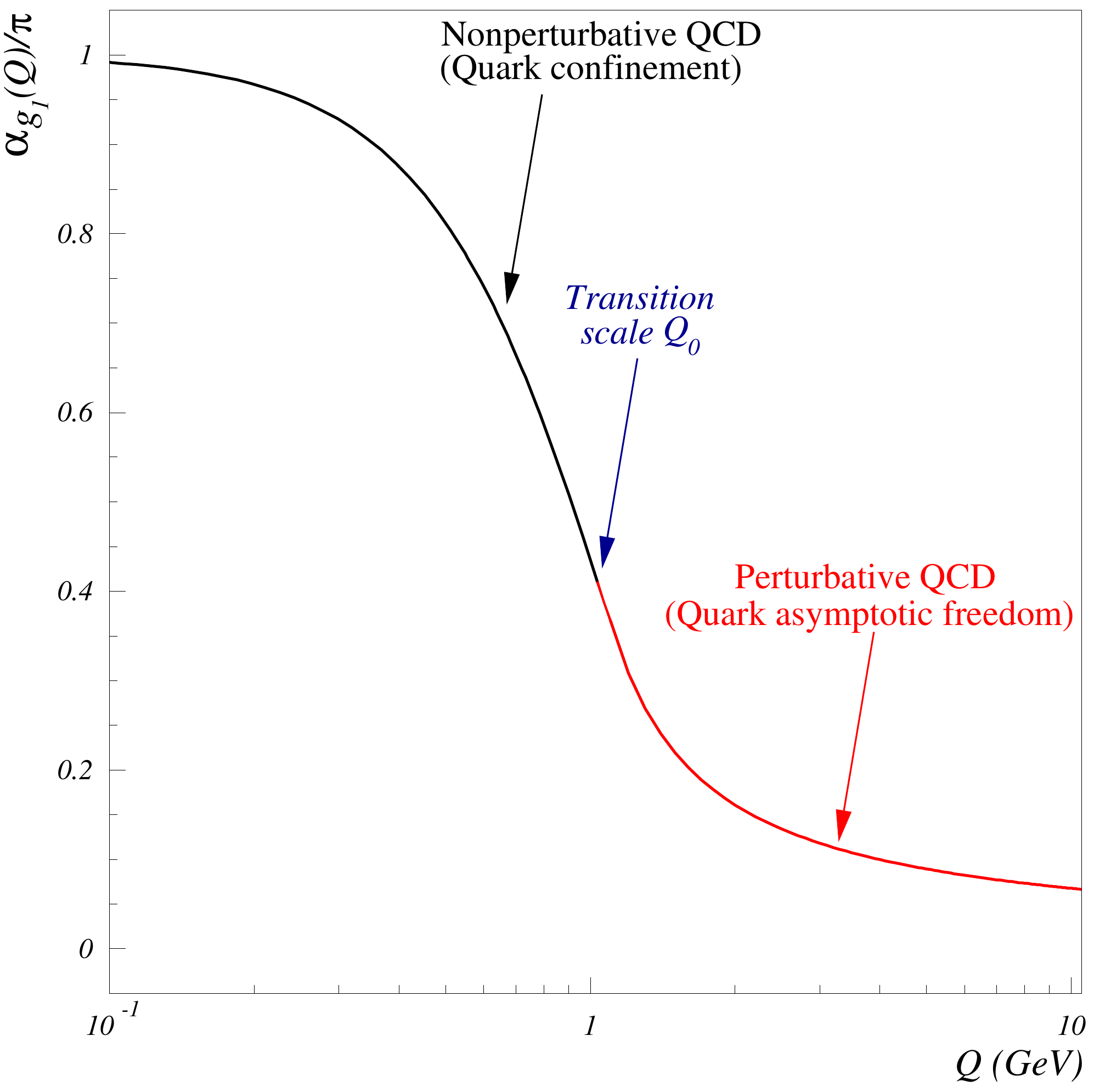}
\caption{\label{Fig:matching}  The strong coupling obtained from the analytic matching of perturbative and non-perturbative QCD regimes.}
\end{figure}

\begin{figure}
\centerline{\includegraphics[width=.42\textwidth]{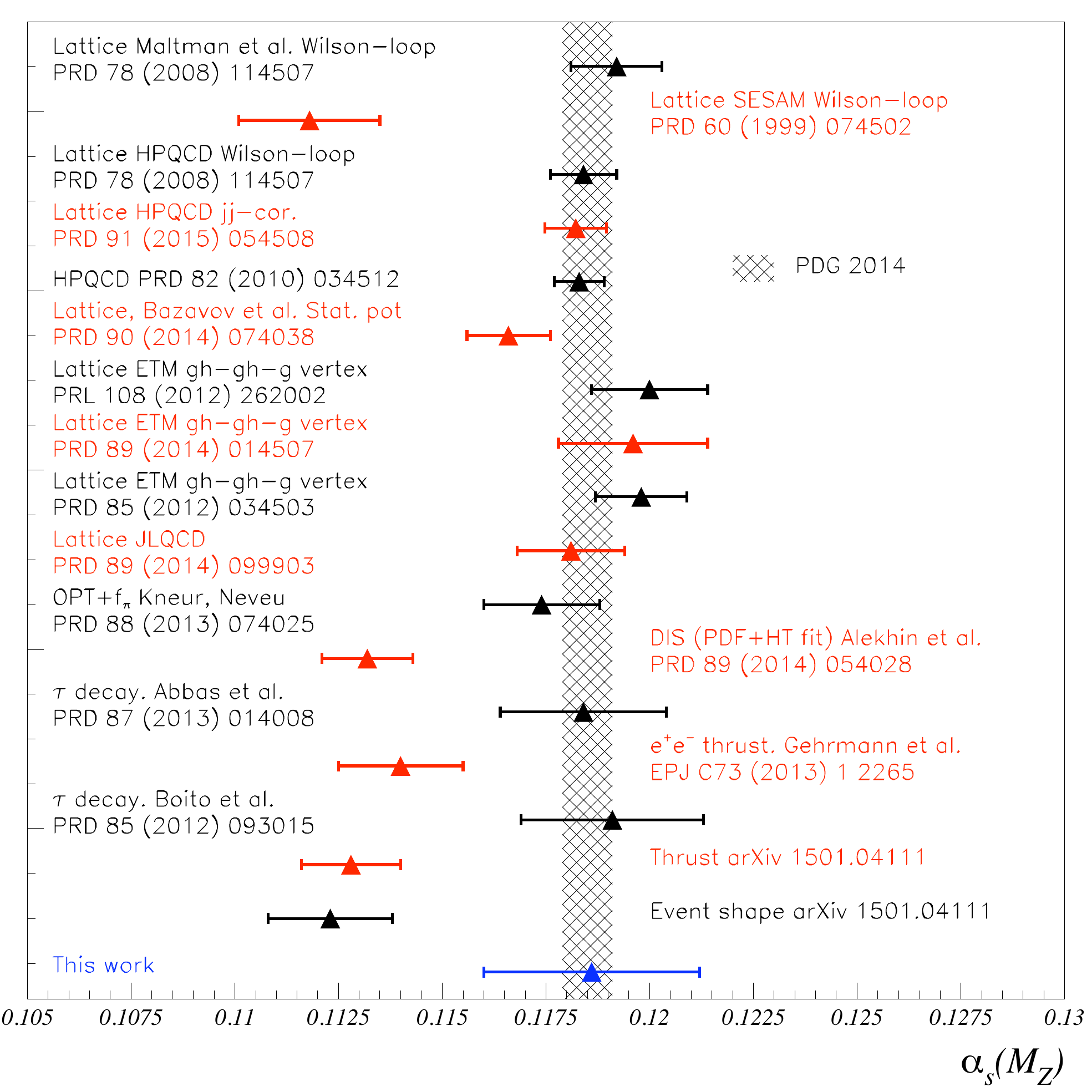}}
\caption{\label{Fig:comparison} Comparison between our result and  determinations of $\alpha_{\overline{MS}}(M_Z)$ from the high precision experimental and lattice measurements. The world average~\cite{PDG:2014} is shown as the vertical band.}
\end{figure}

Eqs.  (\ref{alphaAdS}) and (\ref{eq: alpha_g1 from Bj SR}) thus provide $\alpha_{g_1}(Q^2)$ 
in the large and small distance regimes, respectively.

\section{Relation between $\Lambda_{\overline{MS}}$ and hadron masses}
The existence at moderate values of $Q^2$ of a dual description 
of QCD in terms of either quarks and gluons versus  hadrons (``quark-hadron 
duality"~\cite{Bloom:1970xb}) is consistent with the matching of $\alpha_{g_1}^{pQCD}$ to $\alpha_{g_1}^{AdS}$ at intermediate values of $Q^2$.
This matching can be done by  imposing continuity of both $\alpha_{g_1}(Q^2)$ and their derivatives, 
as shown in Fig. \ref{Fig:matching}. The unique solution for the resulting two equalities determines
$\Lambda_s$ from  $\kappa$, and fixes the scale $Q_0$ 
characterizing the transition between the large and short-distance regimes of QCD.   At leading-order, the 
system can be solved  analytically. It yields:
\begin{equation}  \label{eq: Lambda LO analytical relation}
\Lambda_{\overline{MS}}=M_\rho e^{-(a+1)}/\sqrt{a},
\end{equation} 
with $a=4\big(\sqrt{ln(2)^{2}+1+\beta_{0}/4}-ln(2)\big)/\beta_{0}$. For $n_{f} = 3$ quark flavors, $a\simeq 0.55$. The
system was solved numerically at higher orders. The result at order $\beta_3$, the same order to which the experimental 
value of $\Lambda_{\overline{MS}}$ is extracted, is $\Lambda_{\overline{MS}}=0.341 \pm 0.024$ GeV for $n_{f} = 3$. 
The uncertainty stems from the extraction of $\kappa$  from the $\rho$ or  proton masses and from a small contribution
from  ignoring the quark masses. 

\begin{figure}[h]
\centering
\includegraphics[width=.42\textwidth]{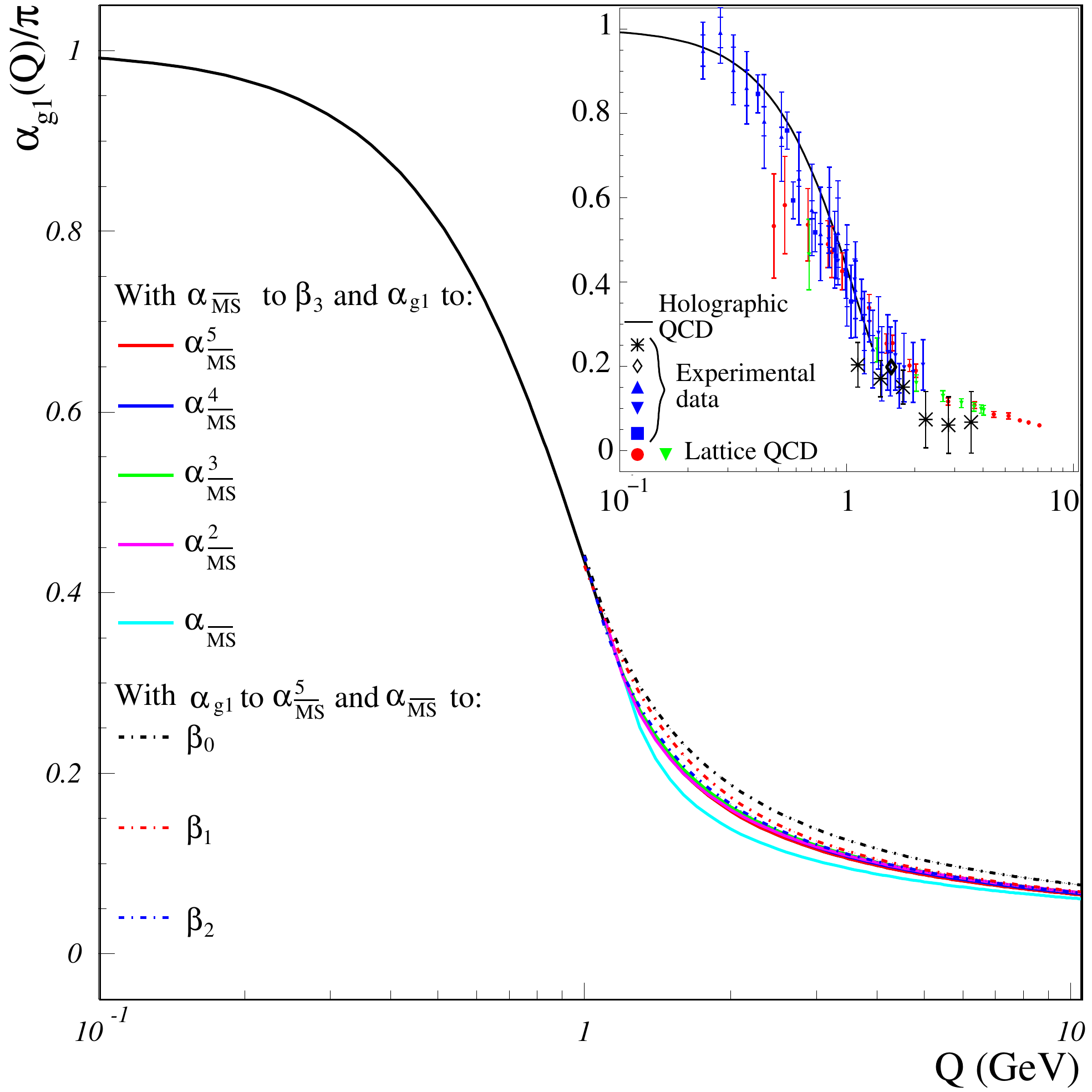}
\caption{\label{Fig:order dependence}  The dependence of $\alpha_{g_1}$ on the orders of the $\beta$ and 
$\alpha_{\overline{MS}}$ 
series.  The continuous black line is $\alpha^{AdS}_{g_1}$. The continuous colored lines are the 
matched $\alpha^{pQCD}_{g_1}$ for all available orders in the $\alpha_{\overline{MS}}$ series (the order of the $\beta$ series 
was kept at $\beta_{3}$). The dash-dotted colored lines are the matched $\alpha^{pQCD}_{g_1}$ at different orders in the 
$\beta$ series (the order of the series was kept at $\alpha_{\overline{MS}}^{5}$).   
 The comparison between $\alpha^{AdS}_{g_1}$ and the data is shown in the embedded figure. 
This comparison is  shown within the  range of validity of AdS/QCD.}
\end{figure}

This theory uncertainty is less or comparable to that of the measurements, which combine to 
$\Lambda_{\overline{MS}} = 0.339 \pm 0.016$ GeV~\cite{PDG:2014}.  In Fig. \ref{Fig:comparison}  we compare 
our calculation with the  best measurements, recent lattice results and their average,  $\Lambda_{\overline{MS}} = 0.340 \pm 0.008$ GeV~\cite{PDG:2014}. 
In  Fig. \ref{Fig:order dependence}, the AdS/QCD prediction of  $\alpha^{AdS}_{g_1}(Q^2)$  
(\ref {alphaAdS}) is plotted together with data~\cite{Deur:2005cf, Deur:2008rf}. Even though 
it has no adjustable parameters, the predicted Gaussian form for the behavior of $\alpha^{AdS}_{g_1}(Q^2)$ at $Q^2 \lesssim 1$ GeV$^2$ agrees well with 
data~\cite{Brodsky:2010ur}. Also shown in this figure is the very small dependence of  
$\alpha^{pQCD}_{g_{1}}(Q^2)$ on the $\beta_{n}$ and  $\alpha_{\overline{MS}}$ 
orders used in Eqs.  (\ref{beta}) and (\ref{eq: alpha_g1 from Bj SR}), respectively.

The matching of the soft and hard domains of the running coupling  $\alpha_{g_1}(Q^2)$ also determines the transition scale $Q_0$. At order $\beta_3$,  
$Q_0^2 \simeq 1.25 \pm 0.19$ GeV$^2$. This value is similar to the traditional lower limit $Q^2 > 1$ GeV$^2$ used for pQCD. 
An approximate value similar to ours was found in Ref.~\cite{Courtoy:2013qca}, which terminates the evolution of 
$\alpha_s(Q^2)$ near $Q \simeq 1$ GeV in order to enforce  quark-hadron 
duality for the proton structure function $F_2(x,Q^2)$  measured in deep-inelastic experiments.
\begin{figure}
\includegraphics[width=.42\textwidth]{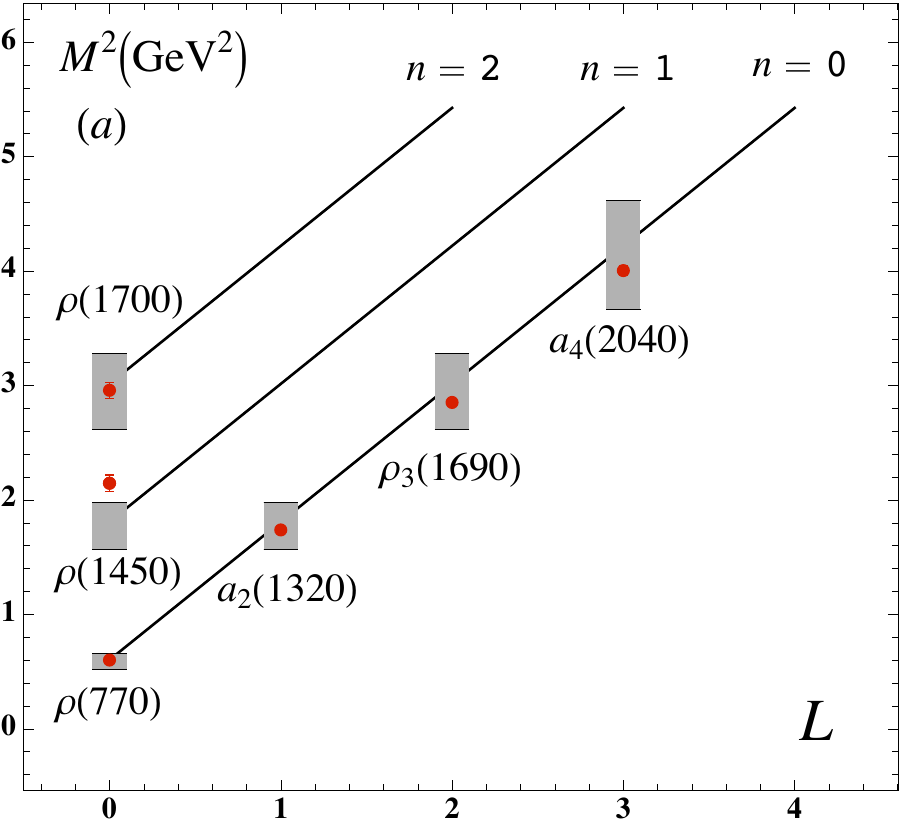}
\caption{\label{Fig:unflavored mesons} The predicted mass spectrum for  the light vector mesons as a function of the internal orbital angular momentum $L$  and the radial excitation $n$ for  unflavored mesons. The red dots are the experimental values. The dark lines represent  the results discussed here and the gray bands the uncertainty. The only parameter entering this determination is the world average $\Lambda_{\overline{MS}} = 0.340 \pm 0.008$ GeV. The decay widths of the mesons are not accounted for in the calculation.}
\end{figure}
\begin{figure}
\includegraphics[width=.42\textwidth]{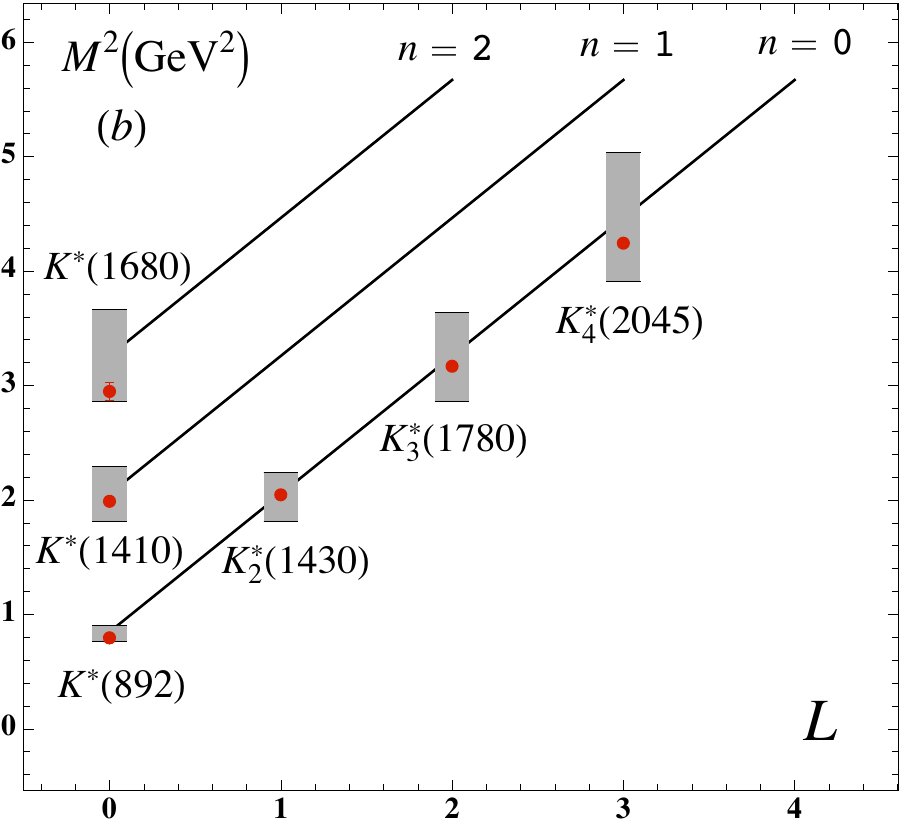}
\caption{\label{Fig:strange mesons} Same as Fig. \ref{Fig:unflavored mesons} but for strange mesons. Only two parameters, 
the strange quark mass and  $\Lambda_{\overline{MS}}$ are used to obtain this spectrum.}
\end{figure}
\section{Determination of the hadron spectrum}
Instead of predicting $\Lambda_{\overline {MS}}$ from $\kappa$, one can, 
conversely, predict the hadron mass spectrum using the world average
$\Lambda_{\overline {MS}}$ = $0.340 \pm 0.008$ GeV~\cite{PDG:2014}
as the only input.   One obtains $M_{\rho}=0.777 \pm  0.051$ GeV, 
in near perfect agreement  with the measured $M_{\rho}=0.775 \pm  0.000$ 
GeV~\cite{PDG:2014}.  The theory uncertainty stems from the truncation of the series, 
Eq. (\ref{eq: alpha_g1 from Bj SR}), from the uncertainty on 
$\Lambda_{\overline {MS}}$~\cite{PDG:2014}, and from the truncation of the 
$\beta$ series, Eq. (\ref{beta}). The computed proton and neutron masses, however, 
are 2$\sigma$ higher than the averaged experimental values, $M_N=1.092 \pm 0.073$ 
GeV compared to $0.939 \pm 0.000$ GeV. Other meson and baryon masses are calculated as 
orbital and radial excitations of the  LF-QCD Schroedinger equation~\cite{Brodsky:2014yha, deTeramond:2013it}.
The  predictions are shown in Figs. \ref{Fig:unflavored mesons} and  \ref{Fig:strange mesons} for the vector mesons. 
Thus, using $\Lambda_{\overline {MS}}$ as the only input, the hadron mass spectrum is calculated 
self-consistently within the holographic QCD framework.

\section{Summary}
We have presented an explicit relation
between the quark-confining nonperturbative dynamics of QCD at  large-distances  and the short-distance dynamics 
of  pQCD; we thus can link the pQCD scale $\Lambda_s$ to the observed hadron masses. The predicted value 
$\Lambda_{\overline{MS}} = 0.341 \pm 0.024$ GeV agrees well with the experimental average 
$0.339 \pm 0.016$ GeV and the lattice average $0.340 \pm 0.008$ GeV.  Conversely, we can predict 
the value of $\kappa$ and the hadron mass spectrum  for light quarks using the experimental value of $\Lambda_s$ as the sole input parameter. 

We  have 
used  an effective theory which encodes the underlying conformality of QCD and the emergence of a scale through  the dAFF procedure. Together with light-front holography, the duality between AdS$_5$ space and physical  $3+1$ space at fixed  LF time $\tau$  allow us to determine both the color confining potential in the Light-Front Schr\"odinger Equation $U(\zeta^2) = \kappa^4 \zeta^2$ and the analytic  form of  the running coupling $\alpha_s(Q^2)/\pi =  \exp{\left(-Q^2/ 4 \kappa^2\right)}$ at small $Q^2.$   The predicted Gaussian form agrees remarkably with the running of  the effective charge determined from measurements of the $Q^2$ dependence of the Bjorken sum rule, in effect, without any free parameters.

It should be emphasized that QCD has no knowledge of conventional units of mass such as GeV; thus only ratios can be predicted from QCD alone. 
The value of $\kappa$ in GeV never needs to be determined.
Consequently our work predicts ratios such as  $\Lambda_s / \kappa$ and $\Lambda_s / M_H$ where  $M_H$ is any hadron mass.  For the 
same reason, only the ratio $\Lambda_{\overline {MS}}/F_{\pi}$ is evaluated in Ref.~\cite{Kneur}.

\paragraph* {\bf Acknowledgments} This material is based upon work supported by the U.S. Department of Energy, Office of Science, Office of Nuclear Physics under contract DE--AC05--06OR23177. This work is also supported by the Department of Energy  contract DE--AC02--76SF00515.

\end{document}